\documentclass[]{IAC_style}

\begin{document}

\IACconference{77} 
\IACpaperyear{2026} 
\IACpapernumber{IAC-26-A4-1-3-x105932} 
\IAClocation{Antalya, Turkiye} 
\IACdate{5-9 October 2026} 

\IACcopyrightA{}

\title{LoFreq-SETI: A Low-Frequency Technosignature Search Pipeline for the LWA Swarm}

\IACauthor{Craig A. Taylor$^{\orcidlink{0009-0006-5879-3105}}$}{a}{1}
\IACauthor{Lillian Ekness$^{\orcidlink{0009-0007-5523-1315}}$}{b}{0}
\IACauthor{Chenoa Tremblay$^{\orcidlink{0000-0002-4409-3515}}$}{{a,c,d,e}}{0}

\IACauthoraffiliation{Department of Physics and Astronomy, University of New Mexico}
\IACauthoraffiliation{Department of Physics, University of Washington}
\IACauthoraffiliation{SETI Institute, Mountain View, CA}
\IACauthoraffiliation{Berkeley SETI Research Center, University of California, Berkeley}
\IACauthoraffiliation{National Radio Astronomy Observatory, Socorro, NM}

\abstract{
The search for extraterrestrial intelligence (SETI) is conducted across the electromagnetic spectrum to understand the prevalence and distribution of intelligent life in the universe. At radio frequencies, this effort focuses on searching for “technosignatures”, observable indicators of technological activity consistent with our understanding of an advanced civilization. Historically, the microwave regime from 1-10 GHz has been the primary frequency range searched for these technosignatures, but recent SETI experiments utilizing low-frequency radio telescopes, such as the Low Frequency Array (LOFAR), Murchison Widefield Array, and MeerKAT radio telescope have explored the insufficiently studied frequency range below 1 GHz. However, the lowest radio frequencies ($<$100 MHz) have been even less explored for technosignatures, despite significant human communication signals occupying this band. The Long Wavelength Array (LWA) is a low-frequency (3-88 MHz) interferometer array telescope collaboration comprised of dipole ‘stations’ distributed across the Southwestern United States, and the LWA Swarm is the aperture synthesis telescope consisting of interconnected LWA stations operating as a long baseline interferometer. Here we present a narrowband technosignature pipeline using the Breakthrough Listen Interesting Signal Search algorithm for raw beamformed single-station LWA data and outline its planned extension to correlated visibilities from the LWA Swarm. We demonstrate the pipeline on a limited selection of fast radio burst observations to evaluate the data processing and spatial coincidence filtering for a future observing campaign of exoplanet systems.  


}


\maketitle
\thispagestyle{fancy} 




\section{Introduction}\label{s:intro}
The search for intelligent life beyond Earth is a fundamental human scientific mission, with efforts spanning the electromagnetic spectrum to understand the prevalence of complex life in the universe. The radio extension of the search for extraterrestrial intelligence (SETI) focuses primarily on discovering intended or unintended emissions consistent with our understanding of purposefully designed technological systems. These signals, called technosignatures, are radio emissions with spectral and temporal structure that are inconsistent with the background of naturally produced electromagnetic signals. Typically, the ideal candidate technosignature would be a narrowband (Hz-scale) beacon with an apparent Doppler drift due to the relative acceleration between observers on Earth and the foreign transmitter. Considerable effort has been invested specifically in the GHz frequency regime, particularly the "water hole" at 1420-1700 MHz \cite{oliver_project_1971,enriquez_breakthrough_2017}, covering large regions of the sky through dedicated and commensal observing systems operating during radio surveys. Low radio frequencies over this same time have not been as thoroughly searched, both in spectral and spatial parameter spaces, due to the unique challenges of observing at MHz frequencies. The combination of terrestrial emissions, the dynamic ionosphere, and the development speed of high-fidelity timing distribution systems has limited the ability for telescopes to achieve sensitivity comparable to that at higher frequencies. Nevertheless, the sole evidence of a technological civilization is derived from humans, who have been utilizing the radio spectrum for communications at frequencies as low as kHz for more than a century. Consequently, it is imperative that we engage in SETI at low radio frequencies.






This description of technosignatures is especially advantageous from a transmission efficiency perspective at meter wavelengths, where radio waves are less affected by line-of sight perturbations (ex. planetary atmospheres, the interstellar medium, dust absorption) and can propagate more freely over large distances at a comparatively lower energy investment than centimetere wave emissions \cite{tremblay_search_2022}. In this effort, a series of SETI efforts from the Murchison Widefield Array \cite{tingay_opportunistic_2016, tingay_search_2018, tremblay_seti_2020, tremblay_search_2022} have utilized both agnostic searches for technosignatures in widefield images and in targeted campaigns of specific sky regions (ex. Galactic Center). These efforts highlight the efficiency of dipole array interferometer-style telescopes in widefield SETI campaigns but also the challenges of band contamination from local sources of radio frequency interference (RFI) and the bright low-frequency radio foreground impacting sensitivity limits. While the radio foregrounds can be modeled, the preponderance of irregular RFI at MHz frequencies is a challenging feature to overcome even with careful characterization of the local interference environment for widefield dipole instruments. One strategy that is employed for astrophysical transient searches is to utilize geographically separated elements and introduce a coincidence filter to preserve only signals detected at multiple telescope elements with matching spectral features. A recent technosignature study utilizing this strategy with two LOFAR stations, one in Ireland and the other in Sweden, found that using dual-station anti-coincidence rejection resulted in all detected narrowband signatures being classified as local RFI \cite{johnson_simultaneous_2023}. 

The Long Wavelength Array (LWA) is an interferometer telescope in the Southwestern United States whose individual elements, called 'stations', are interferometric dipole arrays themselves. Each station is equipped to carry out targeted observations via digital beamforming and form real-time all-sky images as a stand-alone telescope, or conduct joint observations as a long-baseline interferometer collectively called the LWA Swarm \cite{ctaylor2025b}. The frequency coverage of the LWA covers approximately 3-88 MHz, but due to the rapid development of hardware capabilities over the last decade, this range varies slightly from station-to-station. For this project, we utilize the three LWA stations currently operating in New Mexico, the 256-element stations LWA1 \cite{ellingson_lwa1_2013} and LWA-SV \cite{cranmer_bifrost_2017}, and the new cost-efficient 64-element 'swarm' station LWA-NA \cite{ctaylor2025a}. These stations represent the currently connected LWA Swarm telescope and are fully equipped to make coordinated observations as an independent interferometer or through triggered observations with the Karl G. Jansky Very Large Array 4m-band system in the ELWA\footnote{\url{https://github.com/lwa-project/eLWA}} mode (see also EVLA Memos\footnote{\url{https://library.nrao.edu/evla.shtml}} 173, 175, 233). An additional 352-element expanded LWA station is located at the Owens Valley Radio Observatory (OVRO-LWA), but this station is not yet integrated into the LWA Swarm interferometric observing modes, so we have not included it in this work. 

A previous SETI data reduction pipeline for offline processing using the LWA was developed in 2023-2024 using the turboSETI package to look for coincident narrowband signals at two LWA stations, LWA1 and LWA-SV \cite{enriquez_breakthrough_2017, varghese_searching_2024, kulkarni_anti-coincidence_2024}. \texttt{turboSETI} is one of the main Breakthrough Listen narrowband signal search pipelines that has been deployed for regular use in a variety of versions since the survey observations carried out in \cite{enriquez_breakthrough_2017} using the Robert C. Byrd Green Bank Telescope (GBT). It utilizes the Taylor tree method to search dynamic spectra over a set of possible drift rates, set by the frequency resolution and observation duration, by dedrifting the time-frequency diagram at a given drift rate and summing across time to identify frequency spikes in the aggregated dedrifted spectrum. These projects similarly used anti-coincidence to identify unique signals that appeared at similar frequencies at telescope sites separated by approximately 75 km, which were largely attributed to narrowband RFI or unknown sources (S. Varghese; private communication). 

Since this previous technosignature experiment was conducted, a new LWA station was constructed in New Mexico, LWA-NA, allowing for interferometric imaging to be done using the three-baseline LWA Swarm telescope. The Breakthrough Listen collaboration has also developed a new GPU-accelerated technosignature search pipeline to supersede turboSETI, called the Breakthrough Listen Interesting Signal Search (BLISS) algorithm \cite{west_ucberkeleysetibliss_2026}. It is important to note that GPU-enabled variants of turboSETI were implemented after its initial publication (such as seticore\footnote{\url{https://github.com/lacker/seticore}}), but these used the same underlying algorithm to find drifting narrowband signals. BLISS uses a slew of new features such as automated passband equalization, improved noise estimation, flagging using spectral kurtosis and sigma clipping, and desmearing of the drift rate frequency plane when searching for technosignatures. 

This work aims to refresh the original SETI pipeline for the LWA by incorporating a new technosignature search tool, streamline processing for both single-station and visibility data, and perform an automated coincidence signal search on multi-station data. In Section \ref{s:methods} we briefly introduce the methods for data acquisition and preparation for general LWA observing, with subsections \ref{s:methods_pipeline} briefly describing the BLISS toolkit and \ref{s:methods_lofreqseti} outlining the LWA-specific application. The demonstration dataset and results from archival data are provided in Sections \ref{s:val_data} and \ref{s:val_results}. Lastly, Section \ref{s:discuss} discusses the limitations of the raw data that was available for validating the pipeline and further tests needed to better characterize the pipeline. We conclude by laying out a planned exoplanet observing campaign to make a targeted search for technosignatures towards systems with planets in the habitable zone.


\section{Methods}\label{s:methods}

The primary targeted observing mode for the LWA delivers dual-polarization baseband voltage data (51.02 ns minimum cadence) from the digital receiver (DRX) pipelines (commonly called DRX observations) and observations of this type provide two 19.6 MHz tunings around a user-defined center frequency selected from within the station observing band. This baseband voltage product can then be reduced to the user's preference using the tools available in the LWA Software Library (\texttt{lsl}\footnote{\url{https://github.com/lwa-project/lsl}}; \cite{dowell_lsl}). Previous SETI experiments using the LWA were conducted before the development of the LWA Swarm system, and simultaneous observations at different LWA sites had to be manually scheduled \cite{ctaylor2025b, kulkarni_anti-coincidence_2024, varghese_searching_2024}. 

In LWA Swarm interferometry observations, schedules for coordinated observations are distributed to each available LWA station, then the recorded DRX data products are copied to the LWA Users Computing Facility (UCF; \cite{lwamemo_193}). The desired correlation configuration (integration time, FFT-length, polarization basis, etc.) is included in LWA Swarm schedule metadata and forwarded with the DRX data to be correlated by the ELWA correlator running on the LWA UCF. Correlated visibilities are written in FITS-IDI format, delivered to the user for reduction using common radio interferometry calibration software, such as NRAO's Astronomical Imaging Processing System (AIPS) or Common Astronomy Software Applications (CASA), for analysis and imaging. Raw data is retained by the ELWA correlator for 14 days to provide time for provisional quality inspection and to provide an opportunity for users to archive the raw data for reprocessing at a later date. 

The LWA Swarm is currently being redesigned, with new analog and digital hardware to improve bandwidth and sensitivity and reduce the differences in instrumentation and software across the array. To accommodate a new metadata basis with these upgrades and use the most recent tools developed by the Breakthrough Listen collaboration, we designed the LoFreq-SETI\footnote{\url{https://github.com/lillianekness/LoFreq-SETI}} pipeline to conduct interferometric technosignature searches and improve upon the original turboSETI tools. However, due to delays in commissioning the upgrades at LWA and LWA-SV, we were not able to conduct new observations to validate this new pipeline. Instead, we implemented a scheme of patches to make the pipeline backwards compatible with an older LWA data format and validated the tools of this project using archival LWA data\footnote{\url{https://lda10g.alliance.unm.edu/}}. In the following subsection, we describe the main components of the LoFreq-SETI pipeline. 

\subsection{BLISS}\label{s:methods_pipeline}
The Breakthrough Listen Interesting Signal Search (BLISS\footnote{\url{https://github.com/UCBerkeleySETI/bliss}}; \cite{west_ucberkeleysetibliss_2026}) toolkit is the latest GPU-enabled pipeline developed by the international SETI collaboration to hunt for narrowband Doppler drifting signals. This package is intended to supersede the Taylor-Tree de-doppler algorithms of turboSETI and seticore used in previous technosignature programs deployed on a variety of radio telescopes such as the Very Large Array (COSMIC; \cite{tremblay_cosmic_2023, tremblay_cosmics_2025}), Green Bank Telescope \cite{enriquez_breakthrough_2017, margot_search_2023}, and MeerKAT radio telescope (BLUSE; \cite{czech_breakthrough_2021}). BLISS is designed to accept generic time-frequency data from radio telescopes in SIGPROC Filterbank or an equivalent HDF5 format \cite{lorimer_sigproc_2011}. This software is designed to process in units of coarse channels to accommodate the available GPU memory on a given system. This software has previously been published in several works (\cite{sheikh_large-scale_2026}, \cite{li_narrowband_2026}, and \cite{jacobson-bell_breakthrough_2026}), so we will give only a brief description of how BLISS searches spectral data for technosignatures.

The main workhorse of the BLISS package is the \texttt{bliss\_find\_hits} executable that parallelizes a narrowband drifting signal search among incoming polyphase filterbank (PFB) channels, commonly referred to as `coarse channels', that have been finely channelized to some Hz-scale resolution. BLISS preprocesses incoming spectral data to normalize the coarse channels generically to reduce risk of overflowing in 32-bit floating point operations, then applies a passband equalization that can be computed on execution or supplied by the user (commonly by using the BLISS \texttt{gen\_coarse\_channel\_response} function). A series of configurable flaggers are also deployed to help clean out strong RFI spikes in the passband using strategies such as sigma clipping and spectral kurtosis. For each coarse channel, BLISS then searches for local maxima in the Drift Rate -- Frequency plane constructed from the input search parameters, weighting the kernel size of the search by an $L_1$ distance parameter. Maxima in this search window that exceed the desired SNR threshold are then recorded as 'hits' in an unsorted manner to a .capnp file format with the option to output the results as a plaintext .dat file with the following parameters: \texttt{Top\_Hit\_\#, Drift\_Rate, SNR, Uncorrected\_Frequency, Corrected\_Frequency, Index, freq\_start, freq\_end, SEFD (computed System Equivalent Flux Density), SEFD\_freq, Coarse\_Channel\_Number, Full\_number\_of\_hits}. Currently the SEFD and SEFD Frequency fields are currently populated with zero placeholder values and both Uncorrected and Corrected Frequencies are reported using the signal midpoint frequency. Injection and recovery tests have shown that the BLISS narrowband signal search method has better accuracy and precision in recovered SNR and Drift Rate than turboSETI, which searches a range of dedrifted time accumulated 1D spectra, though extensive comparisons have only been conducted on simulated Green Bank Telescope data.

\subsection{The LWA SETI Pipeline (\texttt{LoFreq-SETI})}\label{s:methods_lofreqseti}
\begin{figure}
    \centering
    \includegraphics[width=\linewidth, trim={0 0 0 0}, clip]{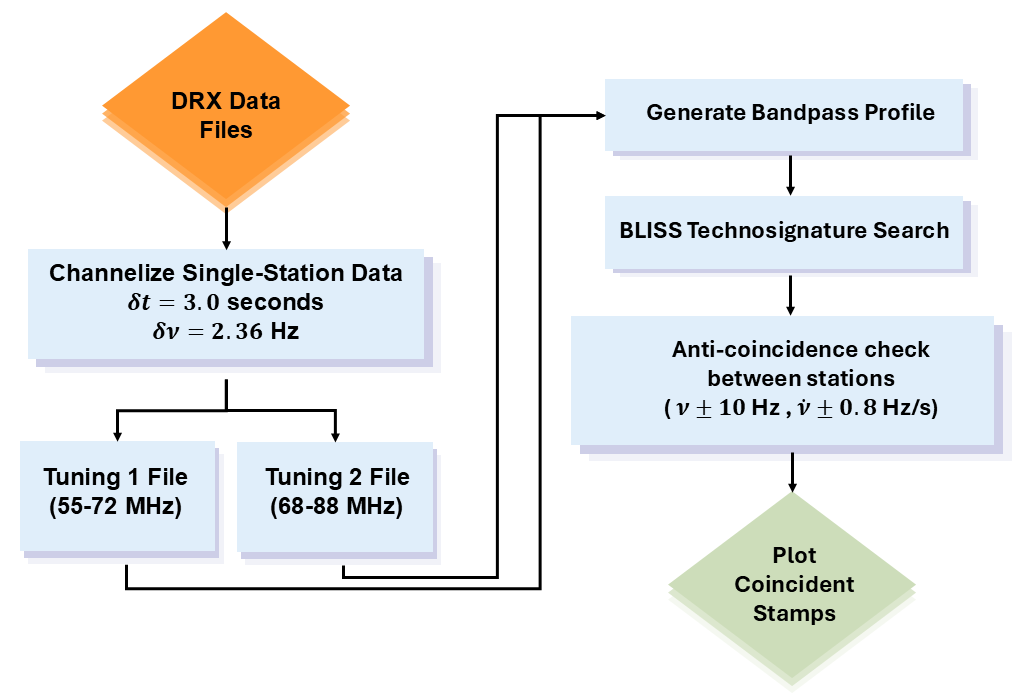}
    \caption{Complete LoFreq-SETI pipeline from raw LWA data through multi-station coincidence plotting for visual comparison. The two halves of the LoFreq-SETI pipeline are denoted by the column structure of the diagram, with the first half preprocessing steps on the left and coincident hit search on the right. The time-frequency resolution used for the validation data and LWA frequency tunings of the observations used herein are listed in the figure for clarity.}
    \label{fig:flowchart}
\end{figure}

The LoFreq-SETI pipeline was developed to translate raw LWA baseband data into an equivalent filterbank-style HDF5 format, utilize the built-in functions of BLISS to search for technosignatures, then perform a coincident signal comparison between simultaneous observations at multiple LWA sites. The pipeline produces a database of technosignature 'hits' detected at each station and combined waterfall plots for similar signals detected at separate LWA stations. These steps are largely divided into two halves for the LoFreq-SETI pipeline, with the first half performing the major data preprocessing steps, and the second half orchestrating the technosignature search with BLISS and a coincident signal search. This package was built by undergraduate student Lillian Ekness during the University of New Mexico 2026 NSF REU program, loosely based on the original turboSETI implementation. The following is a summary of the main steps of the pipeline with a complementary diagram provided in Fig~\ref{fig:flowchart}.

In an observing campaign that we would like to screen for technosignatures, raw LWA DRX observations are recorded at each station using the LWA Swarm observing mode \cite{ctaylor2025b} and copied shortly after observing to the LWA UCF. The first half of the LoFreq-SETI pipeline begins by converting these voltage baseband DRX files into the proper HDF5 format for BLISS in a few steps. This is controlled by a user defined time-frequency resolution to finely channelize the raw data into the standard LWA HDF5 data format for a given search strategy. Next, because LWA frequency tunings are not necessarily continuous or overlapping, we split the data into independent files for each tuning to be processed separately alongside equivalent tuning from other stations. This file division is informed by the metadata associated with each LWA station because the exact analog and digital architecture for each LWA station is slightly different (see \cite{ellingson_lwa1_2013}, \cite{lwamemo_214}, and \cite{ctaylor2025a} for more details). Finally, we convert the file structure into the appropriate HDF5 format for BLISS based on the LWA metadata and this concludes the first half of the pipeline. This is a natural breakpoint in the pipeline because the fine channelization FFT is the largest bottleneck in processing LWA data for technosignature searches, and the remaining components of the full pipeline could be easily broken up into an adapted commensal setting. 

In this HDF5 format, the data has been minimally configured to be accepted by the BLISS software tools and, normally, the built-in functions \texttt{bliss\_generate\_channelizer\_response} and \texttt{equalize\_passband\_filter} would simulate coarse channel equalization then use this model to effectively compensate for PFB effects on incoming spectral data. This is the case for higher frequency radio telescopes that BLISS was primarily considering in its design (ex. GBT, ATA, Meerkat). However, the prevalence of low frequency interference in the observing band of the LWA can result in extremely strong radio frequency interference (RFI) that can affect MHz-scale bandwidths due to spectral leakage. To decrease the impact of individual transmitters contaminating the band, the LWA implements a 4096-channel 4-tap Hamming window polyphase filterbank (PFB) to construct the coarse channels on the full 98 MHz station bandwidth, then a Finite Impulse Response (FIR) filter is applied to only the requested frequency tuning windows\cite{ctaylor2025a}. When we build the fine channelized data using existing \texttt{lsl} tools, these two RFI mitigation steps have already been applied at the closest coarse channel to the user-defined tuning frequency. While the analog filter shape and PFB window can be simulated for removal, a complete inversion of the FIR filter is not readily available.

Since there is not a full analytic model available for all filtering steps, the second half of the pipeline begins by building a tuning dependent composite model to normalize the spectral shape for BLISS analysis. The first component of this `LWA channelizer response' is a frequency-dependent model of the analog signal processor bandpass filter, which is the dominant feature in DRX data and can readily be computed using tools in \texttt{lsl}. Removing this analog feature from the spectrum leaves a residual ripple from the interaction between the PFB coarse channels and FIR filter. This residual is fit using a configurable Savitzky-Golay smoothing filter with localized sigma clipping (default $\sigma_{clip} = 10$) on a median spectrum of a uniform sample of timesteps in a given DRX file. To avoid aliasing at the band edges we assign a new arbitrary coarse channelization regime that is balanced for the computing resources available for running the pipeline, then mask a number of coarse channels at the band edges equivalent to $~10\%$ at each end of the spectra. Each of these two overall bandpass components are normalized, then combined to form the final response model. This process is done on each individual tuning file for each station, then a corrected dataset,  1-D bandpass model, and a manifest of channels affected by sigma clipping are added to the tuning HDF5 files. 

The next step of this pipeline is to pass each tuning file through the \texttt{bliss\_find\_hits} algorithm. The current architecture for pipeline validation leaves a majority of the BLISS settings in their default state, and for the validation dataset we only opt to tune the maximum drift rate search window to constrain the memory usage for benchmarking. The functional call of \texttt{bliss\_find\_hits} searches the incoming unmasked coarse channels for narrowband signals with a $\mbox{SNR}\geq 10$, $L_1$ distance of 7, and over a drift rate range of $\pm3$ Hz/s. Since we have already removed the bandpass profile shape from the spectral data, we supply bliss with a unity equalizer response to avoid \texttt{bliss\_find\_hits} refitting the coarse channels which would taper each coarse channel band edges to reduce aliasing. 

Each LWA station's DRX dataset is passed through the above steps to build a table of narrowband technosignature 'hits' that we cross-match against hits found at other stations (tuning exclusive). This coincidence matching step uses user-defined tolerances in frequency and drift rate to retain matching signals whose reported parameters are approximately equivalent between stations. If a particular signal is found to match at two or more LWA stations, a set of diagnostic plots is produced to show a dynamic spectrum from each site and a window of information describing the coincident hit. An example output coincidence plot is shown in Figure~\ref{fig:hit}. Given that we do not expect a large number of hits to appear simultaneously at many LWA stations separated by tens of kilometers, the pipeline's only output data products are a ledger of coincident hits and their respective multi-station plots for manual inspection and follow-up. If analysis indicates that a coincident hit may be a potential candidate, we could then correlate a small window of the dataset at a similar resolution scale, calibrate the visibilities in AIPS, convert the output into the BLISS HDF5 format, and search the correlated data for the same signal using a similar \texttt{bliss\_find\_hits} configuration. This would greatly improve confidence in the credibility of such a coincident signal, but given the limited resources of the ELWA correlator, this process has not yet been fully integrated into the pipeline. In the future it might also be useful to build a `hits' and `stamps' storage system similar to the COSMIC-VLA data products, but that is beyond the scope of this project \cite{tremblay_cosmic_2023}. 

\section{Demonstration Data and Search Parameters}\label{s:val_data}

From 2021-2023, the LWA conducted a campaign to observe fast radio bursts (FRBs) by accepting triggers from the Canadian Hydrogen Intensity Mapping Experiment (CHIME). In each observation, the raw beamformed voltage data (DRX) from each station was archived for further processing to search for dispersed pulses around the dispersion measure (DM) and timing supplied by the CHIME trigger. At the time of these observations, there were two LWA stations in operation (LWA1, LWA-SV) and of the 129 triggers observed, 95 were concurrently observed at both sites. As described in Section~\ref{s:intro}, the $\sim75~\mbox{km}$ geographic separation of LWA1 and LWA-SV allows the sites to function as a spatial anti-coincidence filter for local sources of RFI, such that we do not expect local RFI to manifest simultaneously at each site. Observation durations range from 2-17 minutes, depending on the dispersion measure reported in CHIME triggers, and were scheduled at overlapping LWA tunings of 53-73 MHz and 68-88 MHz. In total, there are approximately 16 hours of raw data available from both stations. 

Each observation was channelized by the LoFreq-SETI pipeline to a resolution of $\delta t = 3.0~\mbox{s}$ and $\delta f = 2.36~\mbox{Hz}$, and processed on a tuning-specific basis. After searching for hits using BLISS, these observations were searched for multi-station coincidence using a frequency tolerance of $\pm10~\mbox{Hz}$ and a generous range in drift rate of $\pm0.8~\mbox{Hz/s}$ to increase matching signals for validation. This initial validation resolution was selected based on precedent from the previous turboSETI technosignature search, to allow future comparison between the two pipelines, but the pipeline can accept any user defined channelization scheme. However, this frequency resolution scale is well motivated in the context of searching for technosignatures around exoplanets, where the modeled drift rate of a simulated distribution of Kepler stars and confirmed exoplanets place the predicted normalized drift rate ($\dot{\nu}/\nu_{obs}$) ranges at $\pm0.44~\mbox{nHz}$ and $\pm53~\mbox{nHz}$ respectively \cite{li_drift_2022, li_developing_2023}. The drift rate resolution, $\dot{\nu}_{min} = \delta\nu / t_{obs}$, at this channelization for a 180 second observation at 53 MHz is $0.247~\mbox{nHz}$, which could potentially separate the Earth orbit and rotation terms ($\sim0.12~\mbox{nHz}$) from a technosignature around a Kepler-like star (improving for longer observing cadences). The $\pm3~\mbox{Hz/s}$ search window completely encloses the $\pm53~\mbox{nHz}$ known exoplanet drift rate range at the lowest frequency observed, however, the rotational parameters of a large fraction of exoplanets is not well known so this is more effectively a lower limit. This is relevant in the context that this work was originally planned to observe known exoplanet systems in LWA Swarm mode, but the instrument commissioning of upgrades to LWA1 and LWA-SV were not completed at the time of this work. 

To conduct a validation demonstration of the pipeline, we selected 12 CHIME FRB observations that were representative of the different observing durations available distributed over several months (Feb. 2022 -- May 2022). On average, each tuning file per station produced a few hundred hits, but only 8 pairs of hits were found to be coincident across the two stations for the data processed, even considering the generous drift rate tolerance. A summary of all hits as a function of SNR and frequency is shown in Figure~\ref{fig:distr_hits}, where coincident hits between LWA1 and LWA-SV are denoted with stars and a colored sidebar is included to show the spectrum allocations associated with the observing band used. The image products for one such example hit (starred Hit 1 in fig.~\ref{fig:distr_hits}) are provided in Figure~\ref{fig:hit} as an example of the results provided by the LoFreq-SETI pipeline for manual inspection. 


\section{Demonstration Results}\label{s:val_results}
\begin{figure}[htbp]
    \centering
    \includegraphics[width=0.85\linewidth, trim={0 0 0 0 }, clip]{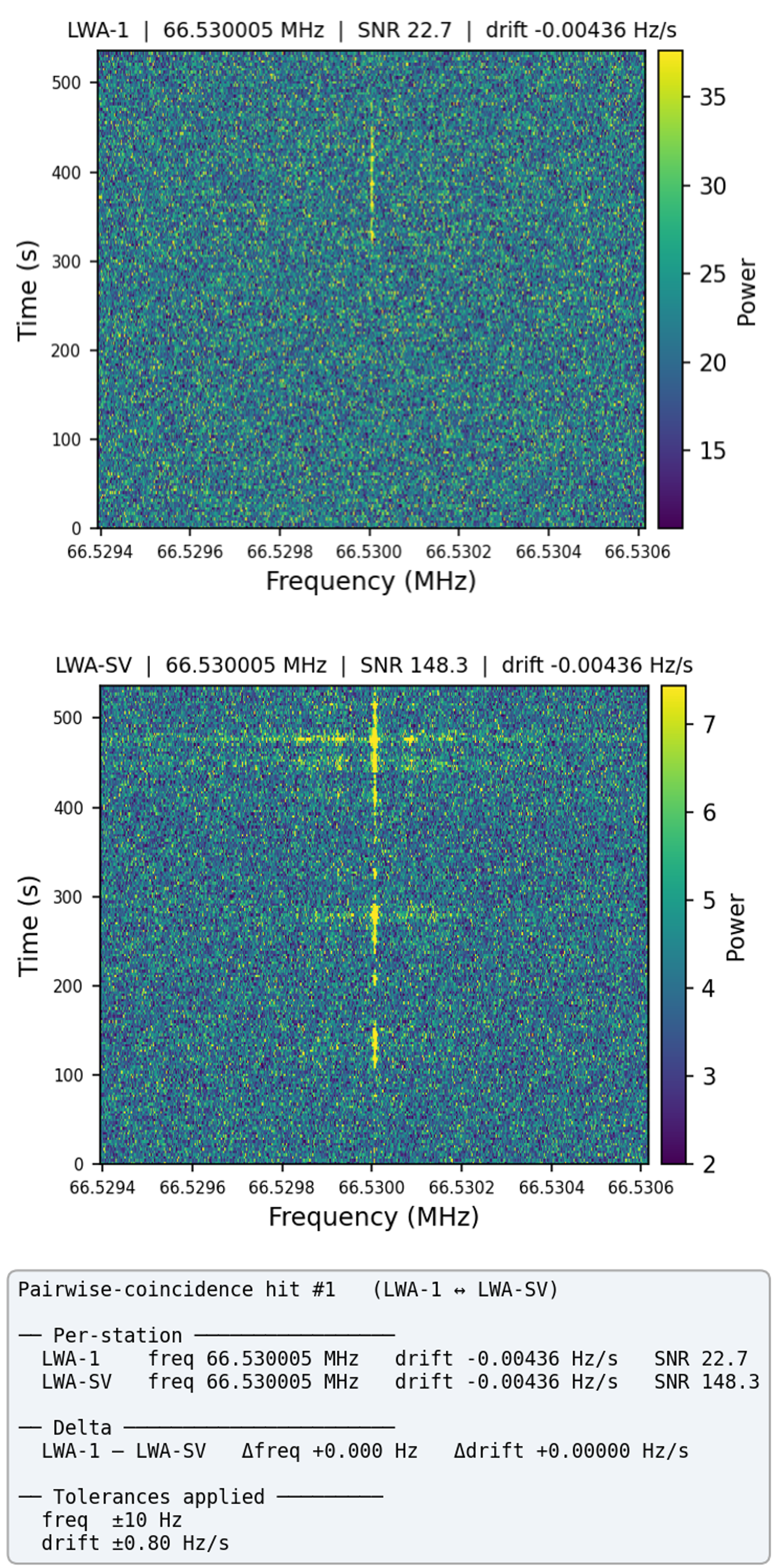}
    \caption{Example multi-station coincident hit output image provided by the LoFreq-SETI pipeline.}
    \label{fig:hit}
\end{figure}

\begin{figure}[htbp]
    \centering
    \includegraphics[width=\linewidth, trim={30 20 50 30 }, clip]{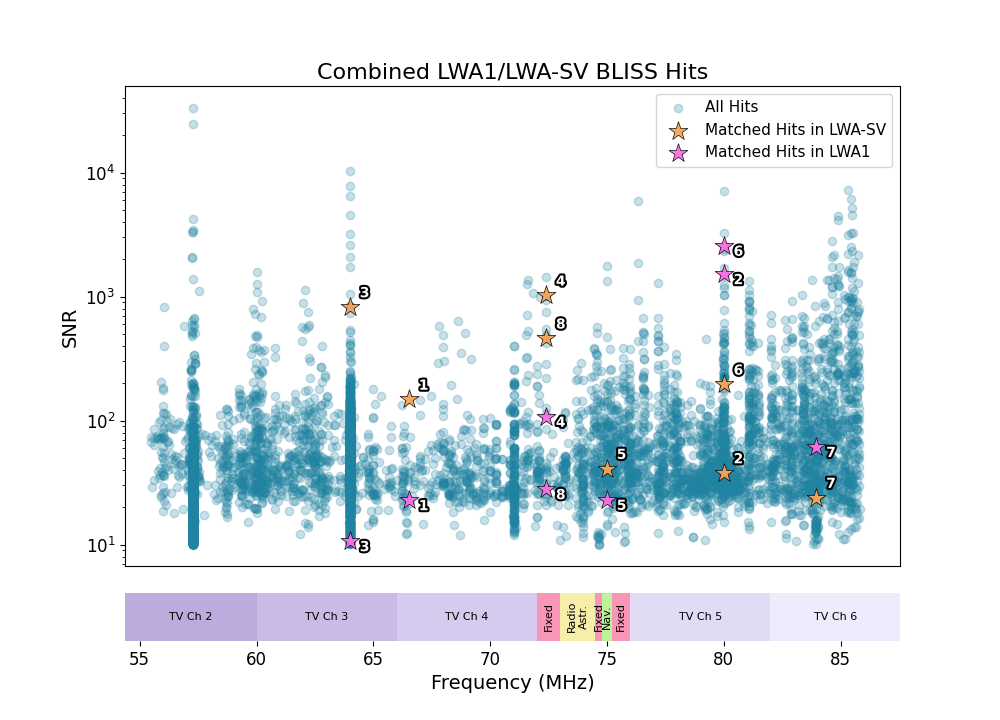}
    \caption{Distribution of hits and coincident hits found in 12 Fast Radio Burst observations taken between Feb-May 2022. The colored bar beneath the plot indicates the National Telecommunications and Information Administration\protect\footnotemark spectrum allocations (Sept. 2025) within the observing band for this project. With the broadcasting broken into individual VHF TV Channels, Fixed indicating Fixed/Mobile allocations, Aeronautical Radionavigation abbreviated to `Nav.', and the VHF Radio Astronomy band.}
    \label{fig:distr_hits}
\end{figure}

In a visual review of the coincident hits from this sample, none of the signal pair spectrograms provided evidence that coincident events could be considered candidate technosignatures and are considered to be due to anthropogenic RFI of some type. For many hits in Fig.~\ref{fig:distr_hits}, including the example provided in Fig.~\ref{fig:hit}, the reported SNR at each station differs at times by more than an order of magnitude. Such a large station-to-station difference in SNR is inconsistent with a cosmic emitter under the approximation that the relative sensitivity of both LWA stations is roughly equal at the time of these observations ($\mbox{SEFD}\approx10~\mbox{kJy}$). We also note the accumulation of hits at specific frequency intervals, including several coincident signals, that are most likely narrowband regions with a high RFI occupancy fraction. 

In six of the coincident hits it is obvious that the narrowband signal detected by BLISS is not drifting at a constant rate, but has obvious curvature that can be periodic in frequency over the duration of a given observation. In single station hits that were not coincident in this case, but could possibly appear with further data reduction of the FRB observations, we also find evidence of emission or reflections from airplanes and satellites in Low Earth Orbit, as the spectrograms exhibit a fixed drift rate for a few minutes before an inflection that inverts the sign of the drift rate. This feature is indicative of a flyby through the primary beam of the telescope, where the source's altitude can be estimated by converting the drift rate into a local orbital acceleration. Again, these cases of drift rate acceleration in the fine spectra at each station are incompatible with cosmic transmissions and are almost certainly local sources of RFI.

\footnotetext{\url{https://www.ntia.gov/page/united-states-frequency-allocation-chart}}

\section{Discussion}\label{s:discuss}

The demonstration presented here establishes that the LoFreq-SETI pipeline is equipped to process LWA beamformed voltage observations into a compatible form for modern technosignature tools, apply BLISS to search those data for narrowband drifting radio signals, and identify signals of interest that appear at approximately equivalent frequencies and drift rates at multiple geographically separated LWA stations. In the 12 FRB observations examined to evaluate this pipeline, individual stations produced several hundred hits per observation, but only a handful of these signals passed the necessary thresholds to be considered coincident at two LWA stations. This reduction in resulting hits that require detailed inspection highlights the utility of spatial anti-coincidence filtering as a discriminatory tool in avoiding sources of RFI.

Of the 8 coincident events that were identified in this limited test dataset, no signal survived visual inspection to be considered plausible technosignatures. Several occurred in heavily occupied bands, which carry a higher possibility of being associated with interference, while 6 signals exhibited non-linear drift characteristics that are not consistent with the constant drift expected for technosignatures observed at this integration interval. Differences in measured SNR between each site in this demonstration oppose a characterization that coincident signals are candidate technosignatures, but highlight the need for more rigorous injection and recovery testing in archival and future LWA data. While this processing demonstrates the successful implementation of an LWA SETI pipeline and the merit of a distributed array in low frequency technosignature studies, we did not identify any strong candidates for astrophysical narrowband emissions in the observations processed. However, a full analysis of the available and suitable archival LWA data has not been completed. 

The work demonstrated here does not make a quantitative analysis of the LoFreq-SETI pipeline for false-positive statistics or provide a complete exploration of the tunable parameters available in BLISS. The archival data available to validate the pipeline required several patches to the original LoFreq-SETI pipeline to accommodate changes to the standard metadata products of the LWA observing architecture. A short commissioning observation including each of the three New Mexico LWA stations was conducted to build the final version of the pipeline, but due to a hardware problem identified during the upgrade, the data at LWA1 and LWA-SV was corrupted by phase instability across the arrays. Nevertheless, the framework of the data was still useful to establish the pathway for future data analysis once suitable repairs can be implemented. 

Many centimeter wavelength SETI campaigns typically use the Kardashev civilization system to describe the sensitivity of their search and frequently reference the Arecibo Planetary Radar 20 TW high frequency transmitter as a benchmark \cite{siemion_1119_2013}. To provide some context in similar terms, we estimate the assumed detectable Equivalent Isotropic Radiated Power (EIRP) for a single LWA station for observations such as those used for pipeline validation. Using the EIRP formula from Johnson \textit{et al.} 2023, for an LWA station with an SEFD of approximately 10 kJy using the observing configuration from this project ($\mbox{SNR}=10$, $t_{obs} = [120,900]~\mbox{s}$, $\delta f = 2.36~\mbox{Hz}$, dual polarization data), this pipeline could be sensitive to a $10^{14}$~W transmitter out to a distance of 10 pc \cite{johnson_simultaneous_2023}. This metric is highly idealized for this LWA implementation as it only considers a zenith beam pointing using SEFD measurements of an older system architecture, so in this context it should be considered an illustrative estimate. In the 12 FRB pointings from this sample, our observed fields contained over 150,000 Gaia stars and 10 confirmed exoplanets from the NASA Exoplanet Archive, highlighting the viability of low frequency instruments to efficiently survey for technosignatures. But a complete reduction for all available FRB observations in the LWA archive and a meaningful direction dependent transmitter limit analysis is required to provide a robust quantitative analysis of LWA technosignature sensitivity limits. 

\section{Conclusions}\label{s:conc}
The LoFreq-SETI pipeline was developed as an NSF REU summer project and this work was organized to update the older pipeline with new software that can be easily deployed for future observing campaigns with the emerging LWA Swarm telescope. Given the duration of the summer program and our inability to record new multi-station observations, we selected a subset of the FRB observations to process for pipeline performance and validation. A future effort will be to process all the FRB observations to perform a comprehensive comparison between this BLISS pipeline and the original turboSETI work. Furthermore, injection and recovery tests could be deployed to better characterize the performance of these technosignature packages for implementation in archival and regular LWA Swarm observing. This is an important next step to understand the differences between turboSETI and BLISS in their performance in processing LWA data for narrowband signals, particularly given the reported differences in SNR calculation and drift rate accuracy reported for the turboSETI package \cite{choza_breakthrough_2023}.

Despite these limitations, we were able to build and test an end-to-end pipeline for reducing single-station LWA data for SETI research using the BLISS toolkit. The pipeline is robust with minimal modifications to handle archival data if it is available in the raw form or, because the LWA archive typically does not save full baseband data, repackaged HDF5 data files with updated headers. With commissioning of the LWA Swarm upgrades expected to be completed by the end of 2026, we plan to continue expanding the SETI capabilities of this deployment on the LWA. Particularly in constructing a robust means to analytically compute the single-station bandpass spectrum to speed up pipeline processing and integrate visibility processing for coincident hits. This pipeline can also be easily configured on the LWA UCF to automatically pick up interferometry observations as they arrive, and process them in tandem with the correlator to search for technosignatures, as is the standard in the field for commensally operated interferometers with added SETI backends. 

In preparation for the REU program, this project was awarded 50 hours of discretionary observing time with the LWA Swarm to observe nearby exoplanet systems to search for technosignatures. This would be conducted in interferometry mode, where we can search both the raw DRX data using the LoFreq-SETI pipeline and modify the BLISS search component to accept correlated visibility data in a suitable format for BLISS. There are no automated calibration pipelines for correlated LWA Swarm observations, so we would first calibrate the data using AIPS to provide a similar data product to that produced by the LoFreq-SETI pipeline. The combination of these datasets in a technosignature search would provide a robust means for identifying and vetting candidate signals, as they would have to pass both an incoherent and coherent search process. Although it is unlikely that many signals will be detected through both these pipeline avenues, it is important nonetheless to continue expanding our search for extraterrestrial intelligence to cover as much of the electromagnetic spectrum as possible. 

\section*{Acknowledgements}
The authors extend thanks to Dr. Savin S. Varghese and undergraduate student Tanvi Kulkarni for building the foundations of this BLISS pipeline using turboSETI, through work as part of the 2024 Berkeley SETI Research Center REU program. 

Construction of the LWA has been supported by the Office of Naval Research under Contract N00014-07-C-0147. Support for operations and continuing development of the LWA is provided by the Air Force Research Laboratory and the National Science Foundation under grants AST-2107845 and AST-2103707.  Renovations of the LWA1 and LWA-SV stations are supported by NASA grant 80NSSC23K0610. This research was sponsored in part by the Air Force Office of Scientific Research (AFOSR) Lab Task 23RVCOR002.


\bibliography{references}


\end{document}